\documentstyle[11pt]{article}
\input epsf
\newwrite\ffile\global\newcount\figno \global\figno=1

\def\writedef#1{}

\def\figin{\epsfcheck\figin}\def\figins{\epsfcheck\figins}
\def\epsfcheck{\ifx\epsfbox\UnDeFiNeD
\message{(NO epsf.tex, FIGURES WILL BE IGNORED)}
\gdef\figin##1{\vskip2in}\gdef\figins##1{\hskip.5in}
\else\message{(FIGURES WILL BE INCLUDED)}%
\gdef\figin##1{##1}\gdef\figins##1{##1}\fi}

\def\figinsert{}
\def\ifig#1#2#3{\xdef#1{fig.~\the\figno}
\writedef{#1\leftbracket fig.\noexpand~\the\figno}%
\figinsert\figin{\centerline{#3}}\medskip\centerline{\vbox{\baselineskip12pt
\advance\hsize by -1truein\center\footnotesize{  Fig.~\the\figno.} #2}}
\bigskip\endinsert\global\advance\figno by1}
\def\endinsert{}

\title{SONOLUMINESCENCE AND THE HEIMLICH EFFECT\footnote{To appear in the
Proceedings of Orbis Scientiae 1996, Miami Beach, FL, January 25-28, 1996.}}

\author{Alan Chodos\\
Department of Physics\\ Yale University, 217 Prospect Street \\
New Haven, Connecticut 06520 -8120\\
chodos@yalph2.physics.yale.edu}
\begin{document}
\setlength{\baselineskip}{24pt}
\maketitle

The phenomenon of sonoluminescence (SL), originally observed some sixty years
ago, $\cite{FS}$ has recently become the focus of renewed interest,
$\cite{GCCR}$-$\cite{FA}$ particularly with the discovery that one can trap a
single bubble and induce it to exhibit SL stably over a large number of
acoustical cycles. $\cite{GCCR}$

In a typical experimental situation, a bubble of gas (usually just air,
possibly doped with a noble gas) in a liquid (usually just water) is made to
expand and then to contract violently under the influence of an applied
acoustic field.  During this motion, $\cite{BP}$ the bubble emits a very sharp
pulse of light, after which it expands again and oscillates about its
equilibrium radius, until stability is regained.  The process then reoccurs in
the next cycle.

Some features to note $\cite{CR}$ are:  (1) the pulse is extremely narrow,
probably not more than 10 ps and possibly much less, whereas the acoustic
frequency is on the order of 30 kHz and the relevant scale for bubble collapse
is perhaps 10 ns; (2) the photon energies are typically at least of order a few
eV, and may be greater.  (The water is opaque to photons with energies beyond 6
eV); (3) The intensity of SL varies considerably depending on a number of
parameters (intensity of the sound field, temperature of the water, composition
of the bubble, etc.) but under optimal conditions pulses with several million
photons are routinely achieved $\cite{HWPB,BWLRP}$;  (4) SL represents a
remarkable concentration of energy: the acoustic energy per atom is typically
eleven or twelve orders of magnitude less than the energies of the individual
photons that are emitted.

On the theoretical side, the problem of understanding SL resolves itself into
three coupled components:  the dynamics of the bubble driven by the sound wave;
the dynamics of the gas within the bubble; and the radiative process that
produces the photons.  The first and second of these $\cite{GCCR,WR}$ would
appear to be classical problems governed by well-known equations (which is not
to say that everything has been understood), whereas the third is undoubtedly a
quantum phenomenon whose origin is still very much in dispute.

In this work we shall adopt a version of the provocative suggestion
$\cite{Schw1}$ put forward by Schwinger:  the mechanism responsible for the
radiation in SL is a dynamic version of the Casimir effect.  It has been known
since Casimir's original work in 1948 $\cite{Cas}$ that the zero-point energy
of quantum fields can be modified by the presence of boundaries, and that these
modifications generate observable effects.  For example, in Casimir's original
work, the quantum fluctuations of the electromagnetic field in the presence of
a pair of uncharged, parallel, perfectly conducting plates were shown to give
rise to an attractive force between the plates.

Schwinger invites us to consider a generalization of the situation, in which
the boundary is that between a dielectric medium (the water) and, essentially,
the vacuum (the gas inside the bubble).  Here, of course, the geometry is
spherical, which already makes the computation more difficult, and an
additional but clearly crucial complication is that the location of the
boundary depends on time.  Under these circumstances, one may expect that
instead of (or perhaps in addition to) the static Casimir force, one will
observe the radiation of the quanta of the electromagnetic field, which will
constitute the sonoluminescence pulse.

The challenge is to present a calculation of this effect that is simple enough
to be tractable, and yet captures the essential physics.  Schwinger struggled
with this problem over the course of seven telegraphic communications
$\cite{Schw1,Schw2}$ to the Proceedings of the National Academy.  Eberlein
$\cite{Eber}$ has done a computation based on an analogy with the Unruh effect
$\cite{Unruh}$, in which the adiabatic approximation is used to permit
quantization in a bubble of fixed radius, and then the photon emission
amplitude is computed to lowest order in the velocity of the bubble.  Her
results illustrate an inherent problem with the dynamic Casimir effect:
Casimir energies tend to be quite small, and in order to reproduce the observed
pulse intensity one must invoke bubble velocities that are rather higher than
seem physically reasonable, even perhaps exceeding that of light.  Milton
$\cite{Milton}$ has done a careful investigation of the static situation and
has pointed out similar difficulties.  Related computations, in a simplified
dynamical model, have been performed by Sassaroli, Srivastava and Widom.
$\cite{SSW}$

In this work we consider a model that neglects the volume effect, i.e. the fact
that light propagates with different velocities in the bubble and in the
medium, and concentrates on the surface effect, i.e. the fact that a boundary
condition must be imposed on the photon field at the surface of the bubble.
Furthermore, we adopt the attitude that, at least for the purposes of an
initial investigation, it should not matter precisely which boundary condition
is chosen.  Thus the choice of a particular form for the term that enforces the
boundary condition will be motivated more by computational convenience than by
an appeal to underlying physical principles, leaving open the future
possibility of finding a more realistic choice.

Thus we consider the action

$$
S = - {1 \over 4} \int d^4x (F_{\mu\nu}F^{\mu\nu} + f(x)
F_{\mu\nu}\tilde{F}^{\mu\nu})  .
$$

Here $\tilde{F}^{\mu\nu} =  {1 \over 2} \epsilon^{\mu\nu\rho\sigma}
F_{\rho\sigma}, F_{\mu\nu} = \partial_{\mu}A_{\nu} - \partial_{\nu}A_{\mu}$ and
$f(x)$ is a dimensionless function that represents the coupling of the photon
to the boundary of the bubble, located at $r = R(t)$, where    $R(t)$ is an
externally prescribed function.   We note that  $F_{\mu\nu}\tilde{F}^{\mu\nu} =
2\partial_{\mu} [\epsilon^{\mu\nu\rho\sigma} A_{\nu}\partial_{\rho}A_{\sigma}]$
so  the  second   term of $S$ may be written ${1 \over 2} \int d^4x
\partial_{\mu} f(x) \epsilon^{\mu\nu\rho\sigma} A_{\nu}\partial_{\rho}
A_{\sigma}$.  If we choose $f(x) = f_0 \theta(R(t) - r)$, we shall obtain a
strictly local coupling of the photon to the surface. Classically $S$ describes
a system obeying the equations of motion

$$
\partial^{\mu}F_{\mu\nu} + \partial^{\mu}f \tilde{F}_{\mu\nu} = 0   .
$$

\noindent\smallskip
which, for our choice of $f$, is solved by a freely propagating electromagnetic
wave subject to the boundary condition

$$
n^{\mu} \tilde{F}_{\mu\nu} = 0
$$

\noindent\smallskip
on the surface, where $n_{\mu}$ is the four-dimensional normal, $n_{\mu} =
{(\dot{R}, \hat{r}) \over \sqrt{1 - \dot{R}^2}}$ .

At the quantum level, the simplest thing to do is to treat the $F\tilde{F}$
term in S as a perturbation, that is, to assume that $f_0$ is a small
parameter.  Whether this is physically reasonable can only be checked a
posteriori, by fitting $f_0$ to the data and seeing if it is indeed small.

In this note, we shall follow this approach, and compute some relevant
amplitudes in lowest order perturbation theory.  Diagramatically, the basic
vertex is

\smallskip
$\left. \right.$  \hspace{-0.6in}\ifig\prtbdiag{}
{\epsfxsize 4truecm\epsfbox{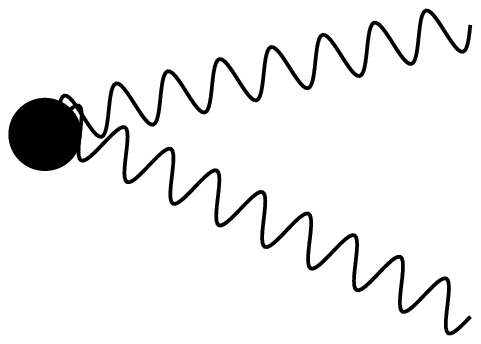}} \vspace{-0.95cm}

\noindent\smallskip
where the dot represents the action of the external source $f(x)$.  It is
actually more convenient to work in momentum space, and therefore we require
the Fourier transform

$$
f(p) =  {1 \over (2\pi)^4} \int d^4x e^{ip\cdot x} f(x)  .
$$

\noindent\smallskip
When $f(x) = f_0 \theta (R(t) - r)$, we find (assuming that $R(t) = R(-t)$,
which is true in Eberlein's model but which does not really fit the data)

$$
f(p_{\mu}) = {-2f_0 \over (2\pi)^3} [{1 \over p} {\partial \over \partial p} Im
{}~g (p, p_0)]  .
$$

\noindent\smallskip
where $g(p, p_0) =  \int_{- \infty}^{\infty} dt e^{ipR(t)} e^{ip_0t}$; and $p =
\mid\vec{p}\mid$.

As a simple example where we shall be able to evaluate $g$ explicitly, we may
consider

$$
R(t) = R_0 ~,  \mid t \mid > T
$$
$$
R(t) = R_0 ~+ v (\mid t \mid - T)  , \mid t  \mid < T  .
$$

\noindent\smallskip
This then yields

$$
Im g = pv \left[ {cos (pR_0 + p_0T) \over p_0 (p_0 + pv)} + {cos (pR_0 - p_0T)
\over p_0 (p_0 - pv)} + {2cos (pR_0 - pvT) \over p^2v^2 - p_0^2}\right]  .
$$

Among the quantities of physical interest that we may compute is the average
number of photons that are produced due to the action of the source:

$$
\left\langle N \right\rangle = \pi^3  \int d^4 P \theta (P_0) \theta(P_0^2 -
P^2) \mid f(P, P_0)\mid^2 (P_0 - P^2)^2   .
$$

\noindent\smallskip
The challenge is to see whether the left-hand side can be of order $10^6$, even
though $f_0^2$, which appears on the right-hand side, is a small number.
Unfortunately, one sees that since $f$ falls off as $p_0^2$ for large $p_0$,
the right-hand side actually diverges.  To ameliorate this, one should include
some or all of the following effects:

(a) instead of a sharp boundary function, $\theta(R(t)-r)$ one should
presumably smooth the boundary over a small distance $\Delta$.  This will
result in a modification of $f$:

$$
f(p, p_0) \rightarrow f(p, p_0) e^{-p\Delta}
$$

\noindent\smallskip
which will not directly solve the large $p_0$ divergence problem for
$\left\langle N \right\rangle$, but will insure that the $p$ integral remains
finite in this and other expressions;

	(b) As Casimir pointed out in his original work $\cite{Cas}$ the boundary that
is represented by $f(x)$ essentially disappears for high frequencies, because
high energy photons do not interact with the boundary as a whole, but only, if
at all, with the individual constituents.  Thus one should insert a high
frequency cutoff on physical grounds;

	(c) The experimental data are cut off by the fact that water absorbs all
photons with energies greater than about 6 eV.  Thus the experimentally
measured $\left\langle N \right\rangle$ is only for photons with energies less
than 6 eV, and hence one should integrate only up to $P_0^2 \sim 12$ eV or so
(the $P_0$ in the integral represents the energy of a pair).
 Of course, one
does not want the water to absorb large numbers of high energy photons:  these
would have observable effects that are not seen.

In addition to the total number of produced photons, it is also possible to
compute other quantities of interest in the same approximation, such as the
spectrum of produced particles.  It seems wise, however, to concentrate first
on $\left\langle N \right\rangle$, since too large an $f_0$ will vitiate the
approximation.  This and related computations are currently in progress.
$\cite{CN}$

We turn next to the consequences of a point we remarked upon earlier:
independent of whatever the radiation mechanism turns out to be, $SL$ can be
viewed as the conversion of acoustic energy, which is distributed diffusely
throughout the liquid, into a burst of electromagnetic energy which is
concentrated spatially into a small region, perhaps a few microns, at the
center of the bubble, and which invests a typical photon with an energy of at
least a few eV, eleven or twelve orders of magnitude more than the acoustic
energy of an atom in the liquid.  Furthermore, this process is reasonably
efficient, in that the energy carried off by the photons is comparable to the
energy required to combat the viscosity of the fluid. $\cite{HPB}$

These features suggest that $SL$ might, with a lot of technological
development, be a candidate for a mechanism of particle acceleration.  After
all, the one indispensable property of an accelerator is not just the energy,
but rather the ability to transfer energy that is macroscopically generated to
individual microscopic particles.  Whatever the mechanism, this is an ability
that $SL$ is observed to possess.

To pursue this idea within the framework of the kind of model that we have been
discussing, we must add to the action the terms, familiar from QED, describing
the electron and its interaction with the electromagnetic field:

$$
S_f = \int d^4 x[\bar{\psi} (i\partial_{\mu}\gamma^{\mu} - m) \psi + e A^{\mu}
\bar{\psi} \gamma_{\mu} \psi]  .
$$

\noindent\smallskip
Then, combining the QED interaction with the photon-bubble interaction, we
shall have diagrams describing the acceleration of the electron.  To lowest
order in the electron-photon interaction, we have

$\left. \right.$  \hspace{-0.6in}\ifig\prtbdiag{}
{\epsfxsize 8truecm\epsfbox{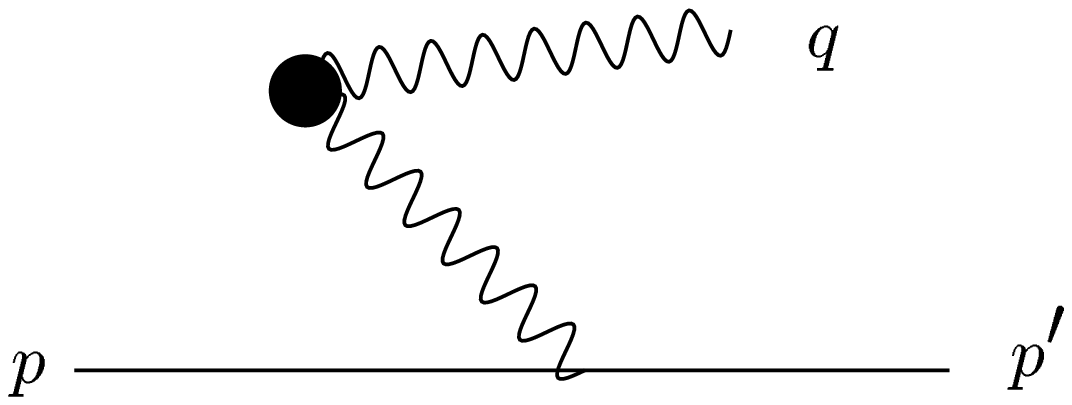}} \vspace{-0.85cm}

\noindent\smallskip
in which the bubble creates a pair and one of the photons then is absorbed by
the electron.  The probability of starting with an electron of momentum $p$
(say, at rest) and measuring an electron of momentum $p^{\prime}$, together
with a photon of momentum $q$, is given by

$$
\overline{\mid \left\langle p^{\prime} q \mid p \right\rangle \mid^2} = {e^2m^2
K(p, p^{\prime}, q) \over (p^{\prime} - p)^4 4\pi q_0 EE^{\prime}} \mid f
(p^{\prime} + q - p) \mid^2
$$

\noindent\smallskip
where the bar indicates an average over the spin of the electron and the
polarization of the photon, and $K$ is a kinematical factor,

$$
K = ({p^{\prime} \cdot p \over m^2} - 1) ((p \cdot q)^2 + (p^{\prime} \cdot
q)^2) - [q \cdot (p^{\prime} - p)]^2    .
$$

\noindent\smallskip
To obtain the total amplitude for acceleration, one should then integrate this
expression over $q$.

We refer to this process as the "Heimlich effect", because, at a rather
different length scale, it produces the same result as the well-known Heimlich
maneuver $\cite{Sted}$:  a bubble is squeezed, and a particle pops out.

We note that the SLC accelerator at SLAC imparts an energy of about 15
eV/micron to the electrons.  Since the size of the sonoluminescent region is of
order a micron, and since the energies are in the 1-10 eV range, we appear to
have the   potential to achieve similar results.  Of course, SLAC is 2 miles
long whereas so far SL has been confined to one micron-sized bubble at a time.
It is premature to speculate on how difficult it might be to improve this
situation.

In summary, we have presented a phenomenological model that we believe captures
the essence of Schwinger's suggestion about the mechanism behind the $SL$
radiation process.  To confront this model with the data, one must achieve
reliable numerical estimates first of $\left\langle N \right\rangle$ and then
of other quantities such as the photon spectrum, not only for the simple $R(t)$
chosen here but for more realistic choices as well.  Furthermore, one should be
prepared to extend the analysis beyond lowest order perturbation theory, and to
take advantage of the quadratic nature of the photon-bubble interaction in
order perhaps to obtain non-perturbative results that will more stringently
test the model. $\cite{CN}$

We have also suggested, regardless of the validity of our model, that the
phenomenon of sonoluminescence may provide the tentative first step toward a
new method of particle acceleration that may be of increasing relevance as the
currently dominant species of accelerator faces inevitable decline and perhaps
even extinction in the twenty-first century.

\end{document}